%% file: taulhh.tex
\documentclass[twocolumn,showpacs,aps,prl,superscriptaddress]{revtex4}

\usepackage{xspace}
\usepackage{graphicx}
\usepackage{dcolumn}
\usepackage{amsmath}
\usepackage{epsfig}

\newcommand{\BABARPubYear}    {05}
\newcommand{\BABARPubNumber} {019}

\newcommand{\SLACPubNumber}{XXXXX}

\input babarsym.tex

\input macros.tex

\newcommand{\lumi}    {221.4\invfb}

\newcommand{\milliontaudecays} {390}
\newcommand{\taulhhlimits} {\ensuremath{(0.7-4.8)\times 10^{-7}}}

\def\figurebox#1#2#3{%
    \def\arg{#3}%
    \ifx\arg\empty
    {\hfill\vbox{\hsize#2\hrule\hbox to #2{\vrule\hfill\vbox to #1{\hsize#2\vfill}\vrule}\hrule}\hfill}%
    \else
    {\hfill\epsfbox{#3}\hfill}%
    \fi}

\begin{document}

\preprint{\babar-PUB-\BABARPubYear/\BABARPubNumber} 
\preprint{SLAC-PUB-\SLACPubNumber} 


\title{
%
{\large \bf \boldmath
Search for Lepton-Flavor and Lepton-Number Violation in the Decay \taulhhs} 
}

\input authors_may2005.tex

\date{\today}

\begin{abstract}
A search for lepton-flavor and lepton-number violation
in the decay of the 
tau lepton into one charged lepton and two charged hadrons 
is performed using \lumi\ of data collected at an 
\epem\ center-of-mass energy of 10.58\gev with the \babar\ detector 
at the \pep2\ storage ring.
In all 14 decay modes considered, the observed data
are compatible with background expectations, and
upper limits are set in the range 
$\BR(\tau\to\ell hh')<\taulhhlimits$ at 90\% confidence level.
\end{abstract}

\pacs{13.35.Dx, 14.60.Fg, 11.30.Hv, 11.30.Fs}

\maketitle


Lepton-flavor violation (LFV) involving charged leptons has 
never been observed, and there are stringent experimental limits 
from muon decays:
$\BR(\mmu\to\electron\gamma) < 1.2 \times 10^{-11}$ \cite{brooks99}
and $\BR(\mmu\to\electron\electron\electron) < 1.0 
\times 10^{-12}$ \cite{sindrum88} at 90\% confidence level (CL).
In tau decays, the most stringent limits on LFV are 
$\BR(\tau\to\mu\gamma) < 6.8 \times 10^{-8}$ and
$\BR(\tau\to\ell\ell\ell) < (1-3)\times 10^{-7}$
at 90\% CL \cite{babar05, babar03}. 
While forbidden in the Standard Model (SM), many extensions to the SM predict enhanced LFV in tau decays with respect to muon decays 
with branching fractions from $10^{-10}$ up to the current experimental limits \cite{ma02}.
Observation of LFV in tau decays would be a 
clear signature of physics beyond the SM, while non-observation
will provide further constraints on theoretical models.

This paper presents the results of a search for lepton-flavor violation in
the neutrinoless decays \taulmhh\ where $\ell$ represents an 
electron or muon and $h$ represents a pion or kaon \cite{cc}.
In addition, a search is also performed for the decays
\taulphh\ which also violate lepton-number conservation.
All possible lepton and hadron combinations consistent with charge
conservation are considered, leading to 14 distinct
decay modes as shown in Table~\ref{tab:results}.
The best existing limits on the branching fractions for these
decay modes currently come from CLEO: $(2-8)\times10^{-6}$
at 90\% CL \cite{cleolhh}.

The data used in this analysis were collected with the \babar\ detector
at the \pep2\ asymmetric-energy $e^+e^-$ storage ring.
The data sample consists of \lumi\ recorded at a
luminosity-weighted center-of-mass energy $\sqrt{s} = 10.58 \gev$.
With an estimated cross section for tau pairs
of $\sigma_{\tau\tau} = (0.89\pm0.02)$ nb \cite{kk},
this data sample contains over \milliontaudecays\ million tau decays.

Charged-particle (track) momenta are measured with a 5-layer
double-sided silicon vertex tracker and a 40-layer drift chamber 
inside a 1.5-T superconducting solenoidal magnet.
An electromagnetic calorimeter (EMC) consisting of 6580 CsI(Tl) 
crystals is used to identify electrons and photons,
a ring-imaging Cherenkov detector (DIRC) and energy loss in the 
tracking system are used to identify charged hadrons, 
and the instrumented magnetic flux return (IFR) is used to
identify muons.
Further details on the \babar\ detector are found in Ref.~\cite{detector}.

A Monte Carlo (MC) simulation of neutrinoless tau decays
is used to study the performance of this analysis.
Simulated \tautau\ events including higher-order radiative
corrections are generated using the \kktwof\ MC generator \cite{kk}, 
with one tau decaying to one lepton and two hadrons 
with a 3-body phase space distribution, while the second tau decay
is simulated with \tauola\ \cite{tauola} according to measured rates \cite{PDG}.
Final state radiative effects are simulated for all decays 
using \photos\ \cite{photos}.
The detector response is simulated with \geant~\cite{geant},
and the simulated events are reconstructed in the same 
manner as data.


Candidate signal events are required
to have a 1-3 topology, where one tau decay yields one
charged particle (1-prong), while the other tau
decay yields three charged particles (3-prong).
Four well reconstructed tracks are required 
with zero net charge, 
originating from a common region consistent with 
$\tau\tau$ production and decay.
Pairs of oppositely charged tracks, likely to be from
photon conversions in the detector material, are ignored
if their \epem invariant mass is less than 30\mevcc.
The event is divided into hemispheres using the
plane perpendicular to the thrust axis, calculated from the 
observed track momenta and EMC energy deposits, 
in the center-of-mass (CM) frame. 
One hemisphere must contain exactly one track while the
other must contain exactly three.

One of the charged particles found in the 3-prong 
hemisphere must be identified as either an electron
or muon candidate.
Electrons are identified using the ratio of observed
EMC energy to track momentum $(E/p)$, the shape 
of the shower in the EMC, and the ionization 
loss in the tracking system $(\dedx)$.
Muons are identified by hits in the IFR
and small energy deposits in the EMC.
Each of the other two charged particles found in the 3-prong
hemisphere must be identified as either a pion or a kaon,
using information from the DIRC and $\dedx$.

After event topology and particle identification requirements,
there are significant backgrounds from light quark 
\qqbar\ production and SM $\tau\tau$ events (without LFV),
as well as small contributions from Bhabha, $\mu^+\mu^-$, 
and two-photon production of four charged particles.
Additional selection criteria, largely the same for all
14 signal channels, are applied as follows.
No photon candidates, identified as EMC energy deposits 
      unassociated to a track, with $E_{\gamma} > 100 \mev$ are allowed.
      This restriction removes \qqbar\ backgrounds
      and SM $\tau\tau$ events.
The total transverse momentum of the event in the CM frame
      must be greater than 0.2 \gevc,
while the polar angle of the missing momentum in the lab frame
      is required to be in the range 
      $[0.25, 2.4]$~radians.
      These two requirements are effective at reducing two-photon
      and Bhabha backgrounds.
The mass of the 1-prong hemisphere
      calculated from the 
      four-momentum of the track in the 1-prong hemisphere and the 
      missing momentum in the event, is required to be in the range
      $[0.6, 1.9]$~\gevcc
      for \ehh\ candidates and 
      $[0.8, 1.9]$~\gevcc for \mhh\ candidates.
      The 1-prong mass requirement is particularly effective at removing 
      \qqbar\ backgrounds as well as the remaining two-photon contribution.
To reduce Bhabha backgrounds, the 
	  momentum of the 1-prong track in the CM frame
	  is required to be less than 4.5 \gevc for the $e\pi\pi$ candidates.
In addition, particle identification vetoes are applied to 
specific selection channels.
For all decay modes, lepton and pion candidates must not pass
the kaon identification as well.
For the \ehh\ decay modes, except for $eKK$, 
the 1-prong track must not be identified as an electron.
This requirement is useful to reduce possible 
contamination from Bhabhas.

To further reduce backgrounds,
candidate signal events are required to have
an invariant mass and total energy in the 3-prong
hemisphere consistent with the neutrinoless decay of a tau lepton.
These quantities are calculated from the observed track momenta 
assuming the corresponding lepton and hadron masses for each decay 
mode.
The mass difference and energy difference are defined as
$\Delta M \equiv M_{\mathrm{rec}} - m_{\tau}$
and 
$\Delta E \equiv E^{\mathrm{CM}}_{\mathrm{rec}} - E^{\mathrm{CM}}_{\mathrm{beam}}$,
where $M_{\mathrm{rec}}$ is the reconstructed 3-prong invariant mass, 
$m_{\tau}=1.777\gevcc$ is the tau mass \cite{bes},
$E^{\mathrm{CM}}_{\mathrm{rec}}$ is the reconstructed 3-prong total energy 
in the CM frame, and $E^{\mathrm{CM}}_{\mathrm{beam}}$
is the CM beam energy.
Rectangular signal regions are defined separately for each 
decay mode in the \dEdM\ plane.
For the \mhh\ modes, \deltaM\ is required
to be in the range $[-20, +20] \mevcc$, while for the \ehh\ modes 
the range is $[-30, +20] \mevcc$ to account for radiative losses.
For all 14 decay modes, \deltaE\
must be in the range $[-100, +50] \mev$.

\begin{figure*}[t]
\resizebox{\textwidth}{!}{%
\includegraphics{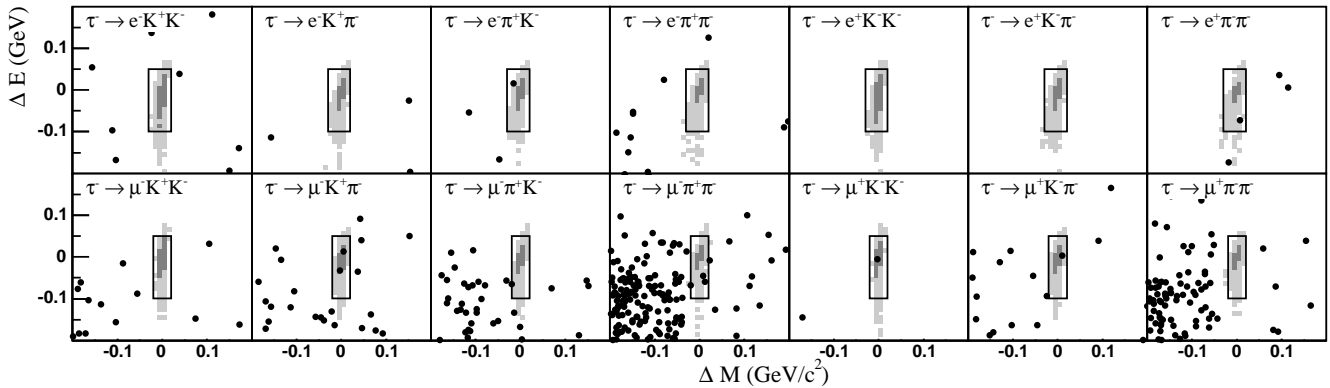}}
\caption{Observed data shown as dots in the \dEdM\ plane and 
the boundaries of the signal region for each decay mode.
The dark and light shading indicates contours containing
50\% and 90\% of the selected MC signal events, respectively.
}
\label{fig1}
\end{figure*}

These signal region boundaries are optimized
to provide the smallest expected upper 
limits on the branching fractions in the background-only 
hypothesis.  These expected upper limits are estimated using 
only MC simulations, not candidate events in data.
To avoid bias, a blind analysis procedure was adopted
with the number of data events in the signal region
remaining unknown until the selection criteria 
were finalized and all systematic studies had been performed.
Fig.~\ref{fig1} shows the observed data for all
14 selection channels,
along with the signal region boundaries 
and the expected signal distributions.

\begin{table}
\begin{center}
\caption{Efficiency estimates, the number of expected background events (\Nbgd) in the signal region (with total uncertainties),
the number of observed events (\Nobs) in the signal region, and the 90\% CL 
upper limit for each decay mode.  
}
\begin{tabular}{lcccc}
\hline\hline
Mode & Efficiency [\%] & \Nbgd  & $N_{obs}$ & UL at 90\% CL\\
\hline
\EKKr &$ 3.77 \pm 0.16 $ & $ 0.22 \pm  0.06 $& 0 &$1.4 \cdot 10^{-7}$\\
\EKPr &$ 3.08 \pm 0.13 $ & $ 0.32 \pm  0.08 $& 0 &$1.7 \cdot 10^{-7}$\\
\EPKr &$ 3.10 \pm 0.13 $ & $ 0.14 \pm  0.06 $& 1 &$3.2 \cdot 10^{-7}$\\
\EPPr &$ 3.30 \pm 0.15 $ & $ 0.81 \pm  0.13 $& 0 &$1.2 \cdot 10^{-7}$\\
\MKKr &$ 2.16 \pm 0.12 $ & $ 0.24 \pm  0.07 $& 0 &$2.5 \cdot 10^{-7}$\\
\MKPr &$ 2.97 \pm 0.16 $ & $ 1.67 \pm  0.29 $& 2 &$3.2 \cdot 10^{-7}$\\
\MPKr &$ 2.87 \pm 0.16 $ & $ 1.04 \pm  0.18 $& 1 &$2.6 \cdot 10^{-7}$\\
\MPPr &$ 3.40 \pm 0.19 $ & $ 2.99 \pm  0.41 $& 3 &$2.9 \cdot 10^{-7}$\\
\EKKw &$ 3.85 \pm 0.16 $ & $ 0.04 \pm  0.04 $& 0 &$1.5 \cdot 10^{-7}$\\
\EKPw &$ 3.19 \pm 0.14 $ & $ 0.16 \pm  0.06 $& 0 &$1.8 \cdot 10^{-7}$\\
\EPPw &$ 3.40 \pm 0.15 $ & $ 0.41 \pm  0.10 $& 1 &$2.7 \cdot 10^{-7}$\\
\MKKw &$ 2.06 \pm 0.11 $ & $ 0.07 \pm  0.10 $& 1 &$4.8 \cdot 10^{-7}$\\
\MKPw &$ 2.85 \pm 0.16 $ & $ 1.54 \pm  0.25 $& 1 &$2.2 \cdot 10^{-7}$\\
\MPPw &$ 3.30 \pm 0.18 $ & $ 1.46 \pm  0.27 $& 0 &$0.7 \cdot 10^{-7}$\\
\hline
\hline
\end{tabular}
\label{tab:results}
\end{center}
\end{table}

The dominant remaining backgrounds are low multiplicity \qqbar{} events
and SM $\tau\tau$ events.
These background classes have unique distributions
in the \dEdM\ plane:
\qqbar{} events populate the plane uniformly,
while $\tau\tau$ backgrounds are restricted to negative 
values of both $\Delta M$ and $\Delta E$.
Backgrounds from Bhabha, $\mu\mu$, and two-photon events are found to be negligible. 
For each background class, a probability density function (PDF)
describing the shape of the background distribution in the \dEdM\
plane is determined by fitting an analytic function to the Monte
Carlo prediction as described in more detail below.
These PDFs are then combined with normalization coefficients 
determined from an unbinned maximum likelihood fit to the observed 
data in the \dEdM\ plane in a sideband (SB) region.
The resulting function describes the event rate observed in the
SB region and is used to predict the expected 
background rate in the signal region.
The SB region is defined as the rectangle, excluding the signal region,
bounding \deltaM\ in the range
$[-0.7, +0.4]\gevcc$ for \ehh\ final states and
$[-0.4, +0.4]\gevcc$ for \mhh\ final states, while
\deltaE\ must be in the range $[-0.7, +0.4]\gev$.
The PDF shape determinations and SB fits are performed
separately for each of the 14 decay modes.

For the \qqbar\ backgrounds, a PDF is
constructed from the product of two functions $P_{M'}$ and 
$P_{E'}$, where the coordinates $(\Delta M', \Delta E')$ 
have been rotated slightly from \dEdM\ to better fit the 
expected distributions.
The function $P_{M'}(\Delta M')$ is a Gaussian and the function
$P_{E'}(\Delta E') = (1-x/\sqrt{1+x^2})(1+a_1 x+a_2 x^2+a_3 x^3)$ where 
$x=(\Delta E'-a_4)/a_5$ and $a_i$ are fit parameters.
The resulting \qqbar\ PDF is described by eight fit parameters,
including the rotation angle, which are determined by fits to MC
\qqbar\ background samples for each decay mode.
For the $\tau\tau$ PDF, the function $P_{M'}(\Delta M')$ is the sum 
of two Gaussians with different widths above and below the peak,
while the functional form of $P_{E'}(\Delta E')$
is the same as the \qqbar\ PDF above. 
To properly model the wedge-shaped kinematic limit in tau decays, 
a coordinate transformation of the form $\Delta M' = 
\cos\beta_1 \Delta M + \sin\beta_1 \Delta E$ and 
$\Delta E' = \cos\beta_2 \Delta E - \sin\beta_2 \Delta M$
is performed.
In total there are 12 free parameters describing this PDF, and
all are determined by fits to MC $\tau\tau$ samples.

\begin{figure}[tbh]
\resizebox{\columnwidth}{!}{
\includegraphics{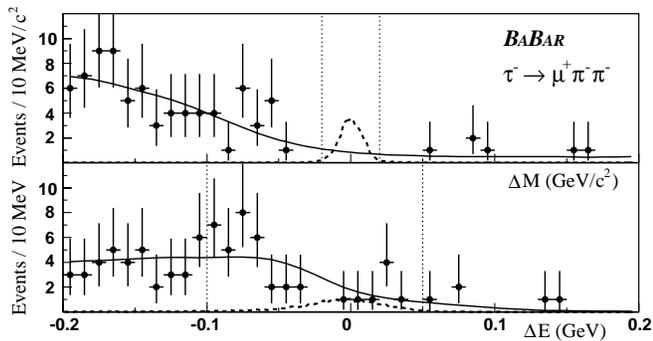}}
\caption{Data (points) and background expectation (solid line)
are shown for the \MPPw\ candidates displayed in Fig.~\ref{fig1}.
Expected signal distributions for a branching fraction of
\mbox{$5\times 10^{-7}$} are also shown as the dashed curve.
The vertical lines indicate the signal region. 
}
\label{fig2}
\end{figure}

With the shapes of the two background PDFs determined, 
an unbinned maximum likelihood fit to the data in the SB 
region is used to find the expected rate of each background 
type in the signal region.
Extensive MC studies show that these PDF functions adequately
describe the predicted background shapes near the signal regions.
The accuracy of these predictions is verified by comparing to
data in regions neighboring the signal region in the \dEdM\ plane
where no signal is expected.
Expected backgrounds are shown in Table~\ref{tab:results},
and an example of the background prediction compared to the observed
data is shown in Fig.~\ref{fig2}.

The efficiency of the selection for signal events is
estimated with a MC simulation of neutrinoless tau decays.
About 40\% of the MC signal events pass the initial 1-3 topology 
requirement, and 20\% to 70\% of these preselected events pass
the particle identification (PID) criteria, depending upon the signal mode.
The final efficiency for signal events to be found
in the signal region after all requirements is shown in 
Table~\ref{tab:results} for
each decay mode and ranges from 2.1\% to 3.8\%.
This efficiency includes the 85\% branching fraction for 1-prong 
tau decays \cite{PDG}.

The PID selection efficiencies and misidentification 
rates are measured directly
using tracks in kinematically-selected data control samples.
These values are parameterized as a function of particle momentum, 
charge, polar angle, and azimuthal angle in the laboratory frame.
The lepton-identification criteria have been designed to give very low
mis-identification rates at the expense of some efficiency loss.
The electron ID is expected to be 81\% efficient in signal 
\ehh\ events, with a mis-ID rate of 0.1\% for pions and 
0.2\% for kaons in generic $\tau\tau$ events.
The muon ID is 44\% efficient for \mhh\ signal events, with
a mis-ID rate of 1.0\% for pions and 0.4\% for kaons.
The hadronic identification is designed to classify the
hadronic candidates as pions or kaons, but is not intended 
to distinguish hadrons from leptons.
The pion ID is 92\% efficient with a mis-ID rate of 12\% for
kaons, while the kaon ID is 81\% efficient with a 1.4\% mis-ID
rate for pions.


The largest systematic uncertainty for
the signal efficiency is the uncertainty in
measuring particle ID efficiencies.
This uncertainty (all uncertainties quoted are relative) 
is dominated by the 
statistical precision of the PID control samples, 
and ranges from $0.7\%$ for \EPPr\ to $3.8\%$ for \MKKr.
The modeling of the tracking efficiency contributes an uncertainty 
of 2.5\%, while the restriction on extra photons leads 
to an additional uncertainty of 2.4\%.
All other sources of uncertainty are found to be 
small, including the modeling of radiative effects, 
track momentum resolution, trigger performance, observables used in 
the selection criteria, and knowledge of the tau 1-prong branching 
fractions.
No uncertainty is assigned for possible model dependence of the signal 
decay.
The selection efficiency is found to be uniform within 
20\% across the Dalitz plane, provided
the invariant mass for any pair of particles 
is less than $1.4 \gevcc$.

Since the background levels are extracted directly from the data,
systematic uncertainties on the background estimation are directly
related to the background normalization, parameterization, and the 
fit technique used.
The finite data available in the SB region used to determine 
the background rates dominates the background uncertainty.
Additional uncertainties of 10\% are estimated by varying the fit 
procedure and changing the functional form of the background PDFs.
The uncertainty on the branching fraction of SM tau decays with
one or two kaons is also evaluated, and contributes less than 15\% 
for all final states.


The numbers of events observed (\Nobs) and the background expectations
(\Nbgd) are shown in Table~\ref{tab:results}, with no significant
excess observed.  Upper limits on the branching
fractions are calculated according to $\BRul = \Nul/(2 \varepsilon \L
\sigma_{\tau\tau})$, where $\Nul$ is the 90\% CL upper limit for the
number of signal events when \Nobs\ events are observed with \Nbgd\
background events expected.  The quantities $\varepsilon$, $\L$, and
$\sigma_{\tau\tau}$ are the selection efficiency, luminosity, and
\tautau{} cross section, respectively.  
The branching fraction upper limits are calculated 
including all uncertainties using the technique of Cousins and 
Highland \cite{cousins92} following the implementation of 
Barlow \cite{barlow02}.  
The estimates of $\L$ and $\sigma_{\tau\tau}$ are correlated \cite{lumi}, 
and the uncertainty on the product $\L \sigma_{\tau\tau}$ is 2.3\%.
The 90\% CL upper limits on the \taulhh\ branching
fractions, shown in Table~\ref{tab:results}, are in the range
\taulhhlimits.  
These limits represent an order of magnitude
improvement over the previous experimental bounds \cite{cleolhh}.

\input acknow_PRL.tex


\end{document}

%% file: authors_may2005.tex
%
\author{B.~Aubert}
\author{R.~Barate}
\author{D.~Boutigny}
\author{F.~Couderc}
\author{Y.~Karyotakis}
\author{J.~P.~Lees}
\author{V.~Poireau}
\author{V.~Tisserand}
\author{A.~Zghiche}
\affiliation{Laboratoire de Physique des Particules, F-74941 Annecy-le-Vieux, France }
\author{E.~Grauges}
\affiliation{IFAE, Universitat Autonoma de Barcelona, E-08193 Bellaterra, Barcelona, Spain }
% ...
\collaboration{The \babar\ Collaboration}
\noaffiliation

%% file: acknow_PRL.tex
We are grateful for the excellent luminosity and machine conditions
provided by our \pep2\ colleagues, 
and for the substantial dedicated effort from
the computing organizations that support \babar.
The collaborating institutions wish to thank 
SLAC for its support and kind hospitality. 
This work is supported by
DOE
and NSF (USA),
NSERC (Canada),
IHEP (China),
CEA and
CNRS-IN2P3
(France),
BMBF and DFG
(Germany),
INFN (Italy),
FOM (The Netherlands),
NFR (Norway),
MIST (Russia), and
PPARC (United Kingdom). 
Individuals have received support from CONACyT (Mexico), A.~P.~Sloan Foundation, 
Research Corporation,
and Alexander von Humboldt Foundation.

%% file: taulhh.bbl
\begin{thebibliography}{99}
\bibitem{cc}
Throughout this paper, charge conjugate decay modes also are implied.
\bibitem{lumi}
The luminosity is measured using the observed \mumu rate, and the \mumu
and \tautau\ cross sections are both estimated with \kktwof.
\end{thebibliography}
